\documentclass[usenatbib]{mn2e}
\usepackage{graphicx}

\newcommand\kms{km$\,$s$^{-1}$}
\newcommand\Msol{M$_{\odot}$}

\begin{document}

\title[Stacking Confusion]{When is Stacking Confusing?: The Impact of Confusion on Stacking in Deep HI Galaxy Surveys}
\author[Jones et al.]{Michael G. Jones$^{1}$\thanks{E-mail: jonesmg@astro.cornell.edu}, Martha P. Haynes$^{1}$, Riccardo Giovanelli$^{1}$ \newauthor and Emmanouil Papastergis$^{2}$
\\
$^{1}$Cornell Center for Astrophysics and Planetary Science, Space Sciences Building, Cornell University, Ithaca, NY 14853, USA
\\
$^{2}$Kapteyn Astronomical Institute, University of Groningen, Landleven 12, Groningen NL-9747AD, Netherlands}

\maketitle

\begin{abstract}
We present an analytic model to predict the HI mass contributed by confused sources to a stacked spectrum in a generic HI survey. Based on the ALFALFA correlation function, this model is in agreement with the estimates of confusion present in stacked Parkes telescope data, and was used to predict how confusion will limit stacking in the deepest SKA-precursor HI surveys. Stacking with LADUMA and DINGO UDEEP data will only be mildly impacted by confusion if their target synthesised beam size of 10 arcsec can be achieved. Any beam size significantly above this will result in stacks that contain a mass in confused sources that is comparable to (or greater than) that which is detectable via stacking, at all redshifts. CHILES' 5 arcsec resolution is more than adequate to prevent confusion influencing stacking of its data, throughout its bandpass range. FAST will be the most impeded by confusion, with HI surveys likely becoming heavily confused much beyond $z=0.1$. The largest uncertainties in our model are the redshift evolution of the HI density of the Universe and the HI correlation function. However, we argue that the two idealised cases we adopt should bracket the true evolution, and the qualitative conclusions are unchanged regardless of the model choice. The profile shape of the signal due to confusion (in the absence of any detection) was also modelled, revealing that it can take the form of a double Gaussian with a narrow and wide component.
\end{abstract}


\section{Introduction}
\label{sec:intro}

The upcoming construction and commissioning of SKA Phase 1 will bring with it a slew of blind HI surveys to be carried out by precursor facilities. While many of these surveys will be shallow or medium depth, wide area surveys, there are several ultra deep single pointing and small field surveys that aim to probe HI galaxies out to unprecedented redshifts.

Stacking has become a key tool for HI astronomers in recent years as measurements of the evolution of HI density with redshift have been attempted \citep{Lah+2007,Delhaize+2013,Rhee+2013}, and low mass and HI-deficient galaxies have been studied at low redshift \citep[e.g.][]{Fabello+2011,Fabello+2011b,Fabello+2012}. As surveys push to increasingly high redshifts, stacking will become an evermore invaluable tool in the attempt to study normal HI galaxies out to a redshift of order unity and beyond.

As surveys become deeper, both in terms of their sensitivity and redshift range, confusion becomes an increasing concern. Longer integration times mean surveys are sensitive to less massive galaxies, but this also means that background emission makes up a larger fraction of the signal detected. Probing HI at higher redshift causes an increasingly large number of objects to be contained in an individual beam width, as the physical size of the beam grows with redshift and therefore encloses more volume. When undetected target objects are stacked this low level emission from the surrounding galaxies will also be coadded. Eventually, when the survey data is deep enough, this confused emission will contribute a significant fraction of the final stacked spectrum and create a bias in the results. The scale of this bias should be estimated so that it can be anticipated and potentially corrected for.

A small number of measurements and predictions of confusion have been made, that are applicable to very deep HI surveys. \citet{Duffy+2008} made predictions for potential FAST (Five hundred metre Aperture Spherical Telescope) surveys, using a similar approach to that used here, but assumed a uniform universe (i.e. neglected the correlation function). As we shall show this leads to an order of magnitude underestimation of the signal due to confusion. \citet{Delhaize+2013} took a different approach by estimating the contribution of confusion in a stack that was known to be heavily confused, based on the optical parameters of the galaxies in the field. This provides a means to interpret a stacked spectrum with confusion, but could also be used to predict the amount of confusion. However, as this would require the specific (optical) input catalogue, and we intend to produce a general tool to assess a generic survey's confusion, this approach will not be discussed in detail in this paper.

In this paper we make use of the currently available HI correlation function (CF) and measurements of the mean ($z=0$) HI density to predict how much HI mass will be contained in a stacked spectrum, in addition to that of the intended targets. This is intended to be a universal tool which can be used to calculate a realistic, but computationally cheap, estimate of the impact of confusion on any HI survey. Section \ref{sec:surveys} briefly outlines the upcoming surveys for which predictions will be made. Section \ref{sec:model} describes how the analytic model is derived, as well as its caveats and limitations. In section \ref{sec:results} we present our results and their implications are discussed in section \ref{sec:discuss}.

\section{Deep Surveys}
\label{sec:surveys}

In the coming years a host of new HI galaxy surveys will begin as part of the precursors to the SKA. In \citet{Jones+2015} we assessed the impact of confusion on shallow and medium depth surveys, whereas this paper focuses on the three deepest of upcoming surveys, LADUMA (Looking At the Distant Universe with MeerKAT), CHILES (COSMOS HI Large Extragalactic Survey) and DINGO UDEEP (Deep Investigation of Neutral Gas Origins - Ultra Deep). We also briefly discuss FAST in a more general sense as the specifics of the surveys it will perform have yet to be determined.

LADUMA \citep{Holwerda+2012} intends to integrate a single pointing with MeerKAT for 5,000 hours. This makes the total field of view of the survey simply the primary beam of a single dish, 0.9 deg$^{2}$ at $z=0$. MeerKAT will have a maximum baseline of 8 km, potentially allowing the synthesised beams to reach down to sizes of $\sim$10 arcsec. The bandwidth of the survey will in theory permit detections of HI sources out to a redshift of $\sim$1.5.

CHILES \citep{Fernandez+2013}, which recently began taking data with the VLA (Very Large Array), is also a single pointing survey, with an integration time of 1,000 hours. Due to the longer baselines of the VLA, the minimum synthesised beam is 5 arcsec across, while the larger dishes reduce the field of view to 0.25 deg$^{2}$ at $z=0$. The narrower bandwidth that CHILES adopts (compared to LADUMA), sets its maximum possible redshift for HI detection at 0.45.

Unlike the two deepest planned pathfinder surveys DINGO UDEEP \citep{Meyer2009,Duffy+2012c} will not be a single pointing. ASKAP (Australian Square Kilometre Array Pathfinder) will survey 60 deg$^{2}$ over the redshift range 0.1-0.43. The survey is intended to be 5,000 hours, and should detect tens of thousands of HI sources. However, due to the computational demands of forming multiple beams (from ASKAP's phased array feeds) and correlating all the signals over this wide bandwidth, it is not yet certain whether ASKAP will achieve a resolution of 10 or 30 arcsec for this survey.

FAST is a single-dish telescope (the only one in this list) currently under construction in China. The 305 m Arecibo observatory in Puerto Rico is the only existing telescope of a comparable size and design. However, unlike Arecibo's fixed reflector, FAST's segmented 500 m primary reflector will be deformable, and the instrument platform will be movable, allowing for zenith angles up to 40$^{\circ}$, which is double the sky area observable from Arecibo. While FAST is observing, a 300 m segment of the reflector will be deformed into a parabola \citep{Nan2006}, giving it a resolution of approximately 3 arcmin for 21 cm radiation, compared to almost 4 arcmin for Arecibo. FAST's larger area will produce greater sensitivity than Arecibo, while its proposed 19 feed horn array (compared to the 7 horn Arecibo L-band Feed Array, or ALFA) will increase its survey speed to a factor of a few faster than Arecibo. Assuming that FAST's feed array has a system temperature of 31 K (as does ALFA) the figure of merit (FoM), which effectively measures a telescope's sensitivity divided by the time taken to map a given area, is 37 for FAST, compared to 4.6 for ALFA on Arecibo (on a scale where 1 pixel with a system temperature of 25 K on Arecibo has a FoM of 1). Although the exact surveys that FAST will carry out have yet to be defined, it has been suggested \citep[e.g.][]{Duffy+2008} that it might probe HI galaxies out to a redshift of $\sim$0.5.

In addition to these upcoming ultra deep surveys, we will reference the two currently available large area, blind HI surveys, ALFALFA (Arecibo Legacy Fast ALFA) and HIPASS (HI Parkes All Sky Survey). The ALFALFA survey \citep{Giovanelli+2005} covers approximately 6,900 deg$^{2}$, with a mean source density of 4 deg$^{-2}$ and a mean redshift of 0.03. The HI properties and functions used throughout this paper \citep{Martin+2010,Papastergis+2013} were derived from the $\alpha$.40 catalogue \citep{Haynes+2011}, which covers 40\% of the nominal sky area. HIPASS \citep{Barnes+2001,Meyer+2004} covers approximately a hemisphere of sky area, but is less deep than ALFALFA, with a mean redshift of 0.01 and a mean source density of 0.2 deg$^{-2}$.

\section{Determining the Confusion in a Stack}
\label{sec:model}

In order to assess how confused a stacked spectrum is, it is necessary to calculate the relative contributions from the target objects versus those they are confused with. The signal due to confusion is found from the total HI mass there is (on average) in a given stack, in addition to that of the target objects. If this mass is negligible in comparison to the mass of the target sources, then clearly it is not a concern. However, if it is comparable in mass, then the spectral profile of this confused emission is also of interest, as this will determine how it alters the appearance of the stacked spectrum in practice. The following subsections outline how each of these quantities can be calculated.

\subsection{Confused Mass in a Stack}
\label{sec:mass_in_stack}

When creating a stack, the angular (or physical) size of the `postage stamps' (or `cut outs') must be chosen. This defines a scale on the sky, and the smallest it can meaningfully be is the size of the beam (or synthesised beam, for interferometers); which is what we shall assume. For simplicity, the fact that the final maps will be made up of pixels is ignored, and the `cut outs' are assumed to be circular. The same analysis could be done with square `cut outs', but given the other uncertainties (see section \ref{sec:lims}) this factor of order unity is unimportant.

Next, a velocity range in the spectrum must be chosen in which the relevant signal is believed to reside. The broadest HI galaxy velocity widths are around 600 km/s, so with an accurate input redshift, a velocity slice of $\pm$300 \kms \ is a conservative choice, and is what will be used here. The results are less sensitive to this choice than might be expected, because the correlation function (CF) causes the signal to be strongly peaked around zero relative velocity.

Together these dimensions define a cylinder in redshift space that is centred on the target being stacked. The amount of HI mass, in addition to the central source, that is within this volume (on average) determines the strength of the confusion signal in the final stacked spectrum, and we will refer to it as the ``confused mass".\footnote{We will also use the phrase ``confused sources" throughout this paper to mean the sources that a target object is confused with, not including the target object itself.} 

In order to calculate the mean confused mass in a stack, two things must be known: the expected number of HI galaxies residing in the cylinder surrounding the target object, and the mean HI mass of an HI-selected galaxy. The first of these can be calculated from the CF, and the second by the integral of the HI mass function (HIMF).

The CF is the excess probability (above random) of two sources being separated by a given distance, here denoted by $\xi(\kappa,\beta)$, where $\kappa$ is the separation perpendicular to the line of sight, and $\beta$ is the separation along it. In general it is not symmetric with respect to $\kappa$ and $\beta$, as distance along the line of sight is usually determined from redshifts, and so peculiar velocities alter the derived separations. Although these distortions along the line of sight are not physical, in the sense that the galaxies may not be separated by the distances calculated, they are directly applicable to this scenario as the depth of the cylinder is also a pseudo-distance (a velocity divided by the Hubble constant). Thus, we make use of the 2D CF for HI sources, as calculated in \citet{Papastergis+2013}, and for convenience, will use the simple analytic fit from \citet{Jones+2015} to approximate it:
\begin{equation}
\xi(\kappa,\beta) = \left( \frac{1}{r_{0}}\sqrt{\frac{\kappa^{2}}{a^{2}} + \frac{\beta^{2}}{b^{2}}} \right)^{\gamma},
\label{eqn:2dcf}
\end{equation}
where $ab = 1$, $r_{0} = 9.05$ Mpc, $a = 0.641$, and $\gamma = -1.13$.

Integrating $1+\xi$ over the cylinder defined by the choice of `postage stamp' size and velocity range, and multiplying by the mean HI source number density, gives the expected number of additional HI sources within the volume. Finally, multiplying by the mean HI mass of an HI source \citep{Martin+2010}, returns the total mass in these sources within the beam on average, $M_{\mathrm{conf}}$.
\begin{eqnarray}\nonumber
M_{\mathrm{conf}} & = & 4 \mathrm{\pi} \Omega_{\mathrm{HI}} \rho_{\mathrm{c}} \int_{0}^{\beta_{\mathrm{sep}}} \int_{0}^{\kappa_{\mathrm{sep}}} \kappa \left[ 1+\xi(\kappa,\beta) \right] \mathrm{d}\kappa \mathrm{d}\beta \\ 
& = & 2 \mathrm{\pi} \Omega_{\mathrm{HI}} \rho_{\mathrm{c}} a \left[ \frac{\beta_{\mathrm{sep}} \kappa^{2}_{\mathrm{sep}}}{b a^{2}} + I \right],
\label{eqn:model}
\end{eqnarray}
where
\begin{eqnarray}\nonumber
I &=& \frac{2\frac{\beta_{\mathrm{sep}}}{b} \left( \frac{\kappa_{\mathrm{sep}}}{a} \right)^{\gamma+2}(\gamma+3)}{(\gamma+2)(\gamma+3) r_{0}^{\gamma}} \\ &&\left[ _{2}F_{1}\left( \frac{1}{2},-\frac{\gamma}{2}-1;\frac{3}{2};-\frac{a^{2} \beta^{2}_{\mathrm{sep}}}{b^{2} \kappa^{2}_{\mathrm{sep}}} \right) -2\left( \frac{\beta_{\mathrm{sep}}}{b} \right)^{\gamma+3} \right]
\end{eqnarray}
and $\mathrm{_{2}F_{1}}$ is the Gaussian hypergeometric function, $\beta_{\mathrm{sep}}$ is the velocity half range, in this case $300/70$ Mpc, $\kappa_{\mathrm{sep}}$ is the physical radius of the beam in Mpc at the distance of the target object, and $\Omega_{\mathrm{HI}}$ is the background density of HI ($\rho_{\mathrm{HI}}$) relative to the critical density ($\rho_{\mathrm{c}}$) in \Msol$\,\mathrm{Mpc}^{-3}$ (equivalent to the mean source number density times the mean source mass). We adopt $\Omega_{\mathrm{HI}} = 4.3 \times 10^{-3}$, as found by \citet{Martin+2010}. Refer to \citet{Jones+2015} for the full details of the fit to $\xi(\kappa,\beta)$ and how to evaluate its integral.

The above equation for the confused mass is independent of the shape of the HIMF, because the quantity of HI in a given volume only depends on its integral. However the variance of the confused mass is dependent on the shape of the HIMF. This can be understood by considering where most of the HI mass in the Universe resides, which at present is in $M_{*}$ galaxies. If the faint-end slope was steeper and most of the HI mass resided in highly abundant dwarf galaxies, then the variance in the confused mass would be small (ignoring the environmental dependence that would likely be present in such a universe) as the Poisson noise in the number counts within the cylinder would be low. Alternatively if the faint-end slope were to be very flat and the knee mass very high, then although the integral could be identical, most of the HI mass would be contained in exceptionally rare, highly massive systems. As a result the Poisson noise associated with the counts of such galaxies would be very large, leading to high variance in the confused mass.

\subsection{Spectral Profile of Confusion}
\label{sec:spec_prof}
\begin{figure}
\centering
\includegraphics[width=\columnwidth]{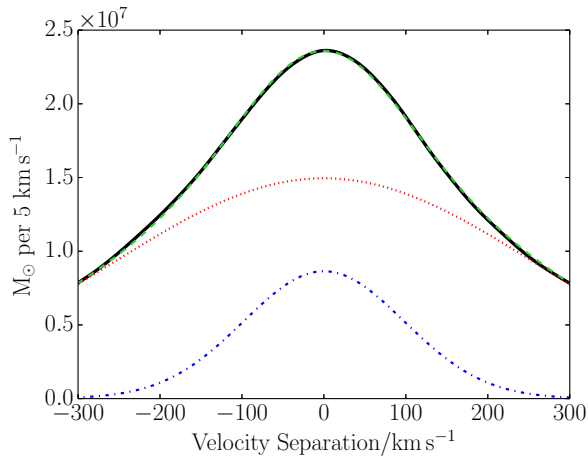}
\caption{The solid black line shows a simulated stack of the average spectral profile contributed by confused sources only (target sources have been removed), in a stack at $z=0.029$ for a survey with a beam size of 15.5 arcmin (at $z=0$), intended to mimic the \citet{Delhaize+2013} experiment with HIPASS. The green dashed line shows the double Gaussian fit to the black profile, while the red dotted and blue dash-dot lines show the two separate components of the fit.}
\label{fig:spec_prof}
\end{figure}
If all the additional mass in the cylinder was uniformly distributed in velocity space then it would not pose a problem to deriving physical properties from the stacked spectrum, as the confusion signal would just represent a DC shift in the baseline. However, if the confusion signal is peaked around the central frequency, then it can contribute an unknown amount to the final flux, or worse, make up all of the flux and give a false positive (in the event that the central sources are not detected even in the stacked spectrum).

The spectral shape of the confusion signal (which we will refer to as the ``confusion profile") can be calculated using a similar method to that in section \ref{sec:mass_in_stack}, which reveals it takes a double Gaussian form. However, this method neglects the velocity widths of each galaxy contributing to the confusion signal. Therefore, we have estimated the confusion profile using mock stacks (see section \ref{sec:stack_sims}), shown by the solid black line in figure \ref{fig:spec_prof}. The inclusion of velocity widths broadens the confusion profile, however it maintains a double Gaussian shape (see figure \ref{fig:spec_prof}). The two components arise from the peak in the CF at zero velocity separation, and the uncertainty in the input catalogue of target redshifts. For the latter we assume a Gaussian distribution centred on zero with a width of 35 \kms, as found by \citet{Toribio+2011}.

Here it should be reiterated that figure \ref{fig:spec_prof} includes only the stacked emission of the confused sources, with emission from the target galaxies removed. The profile is well fit by a double Gaussian with a narrow and a broad component, which highlights that caution must be used when interpreting heavily confused stacks, as this profile shape is similar to what might be expected for a stack detection on top of confusion noise, not just from confusion alone.

\subsection{Mock Stacks}
\label{sec:stack_sims}

In order to help assess our findings and potential strategies to mitigate confusion, we make use of simulated HI stacks. Our approach is similar to that of \citet{Maddox+2013}, which used the template HI profiles of \citet{Saintonge+2007}, however our mock stacks are intentionally noiseless and the masses and velocity widths are drawn randomly from a fit to the $\alpha$.40 mass-width function \citep{Papastergis+2015,Jones+2015}. 

When simulating the signal from confusion, galaxy masses and widths are drawn from the mass-width function. A lower HI mass bound (of $10^{6.2}$ \Msol, the lowest that ALFALFA can measure the HIMF to) must be set, and only masses greater than this are selected. However, the results are insensitive to this bound as most of the HI mass in the Universe is contained in much more massive systems. The number of confused sources to be included (around each target object) is chosen from a Poisson distribution with an expectation equal to $M_{\mathrm{conf}}/\bar{M}_{\mathrm{HI}}$, where $\bar{M}_{\mathrm{HI}}$ is the mean HI mass of a galaxy, and $M_{\mathrm{conf}}$ is the confused mass as calculated in equation \ref{eqn:model}. The galaxy masses and widths are then drawn from the mass-width function and are placed at angular and velocity separations (away from the central target) drawn from the 2D CF (equation \ref{eqn:2dcf}). Finally, the profiles are added to the stack at the appropriate frequencies (the angular information is ignored except when non-uniform beam weightings are consider in section \ref{sec:beam_weight}).

To simulate the contribution of the target objects, we make the assumption that all the targets have the same mass and then draw only the velocity width (for the relevant mass) from the mass-width function. A redshift error is added to the profile, drawn from a Gaussian of width 35 \kms, and then it is added to the stacked spectrum. All stacked targets are assumed to be the same mass for simplicity and generality, however information about the mass distribution of targets, which might be available when modelling a particular survey, would be straightforward to incorporate. This assumption has no impact on the amount of confused mass we calculate, but could alter ratio of confusion to target signals.

\subsection{Modelling Limitations}
\label{sec:lims}

The model and simulation methods described above have a number of caveats and shortcomings which are outlined in this section. A general note is that this methodology only applies to the average values present in a large stack. This will require on the order of 1,000 spectra in a given stack, such that extreme cases and small number statistics are not dominant.

\subsubsection{Redshift Evolution}
\label{sec:red_evo}

Although there is some evidence for $z$-dependence of $\Omega_{\mathrm{HI}}$ from stacking, damped Lyman-$\alpha$ observations and HI intensity mapping experiments \citep[e.g.][]{Rao+2006,Lah+2007,Prochaska+Wolfe2009,Chang+2010,Freudling+2011,Delhaize+2013,Rhee+2013,Hoppmann+2015}, there is no observational data describing how the shape of the HIMF may evolve, or how the HI CF evolves. Due to these limitations we choose to display our results for two separate assumptions: constant $\Omega_{\mathrm{HI}}$, and $\rho_{\mathrm{HI}} \propto (1+z)^{3}$, with both using the $z=0$ CF throughout. The first case will likely under predict the confused mass at high redshift as the observations indicate a factor of $\sim$2 increase in $\Omega_{\mathrm{HI}}$ by $z=1$, while the second case actually appears to overestimate the increase of HI density with redshift. Thus, barring a major shift in the HI CF, we expect the true value to lie between these two cases.

\subsubsection{Sharp Edges \& Point Sources}
\label{sec:edges_points}

The response of the telescope beam is assumed to be a step-function. When stacking based on `cut outs' from a uniform survey map, where the shape of the beam response has already been accounted for, this is the simplest choice. In section \ref{sec:beam_weight} we discuss the possibility of using a different weighting as a way to reduce confusion.

When simulating stacks to verify the analytic results and test mitigation strategies (section \ref{sec:mit_strat}), the confused sources, which in reality would be galaxies with their own velocity widths and spatial patterns, are modelled with realistic HI profile shapes \citep{Saintonge+2007} in frequency space, but as point sources on the sky. Given the simplistic weighting of the beam, modelling sources as points (spatially) is sufficient. However, as the finite velocity widths inevitably broaden the profile of any confusion signal (see figure \ref{fig:spec_prof}), it might be expected that the total mass within a $\pm$300 \kms \ window might differ from the value derived via equation \ref{eqn:model}. This has been explicitly checked for in our mock stacks, and while the spectral profile of the confusion signal becomes broader, it maintains a double Gaussian shape and the total confused mass is consistent with the analytic model.

\subsubsection{Redshift Error Distribution}

In order to stack non-detections an input (presumably) optical catalogue of positions and redshifts must be used. When calculating the profile of the confusion signal a Gaussian distribution with a width of 35 \kms \ is assumed to represent the deviations between the HI and optical redshifts. In practice the scale of this dispersion is dependent on the quality of the spectra in the input catalogue. \citet{Maddox+2013} found a smaller dispersion between SDSS and ALFALFA when only including the highest S/N ALFALFA detections, while \citet{Delhaize+2013} quoted the uncertainty in their input redshifts as 85 \kms. Although the value we chose to adopt changes the width of our resulting profile, it does not alter the qualitative results.

For a particular survey there may be more knowledge about how these redshifts differ from each other which, when available, should be used instead. Alternatively, the bias from confusion could be estimated by calculating the cross correlation function between the HI and optical catalogue when possible, and use this in place of equation \ref{eqn:2dcf}.

\subsubsection{Model Uncertainties and Variance}
\label{sec:mod_var}
\begin{figure}
\centering
\includegraphics[width=\columnwidth]{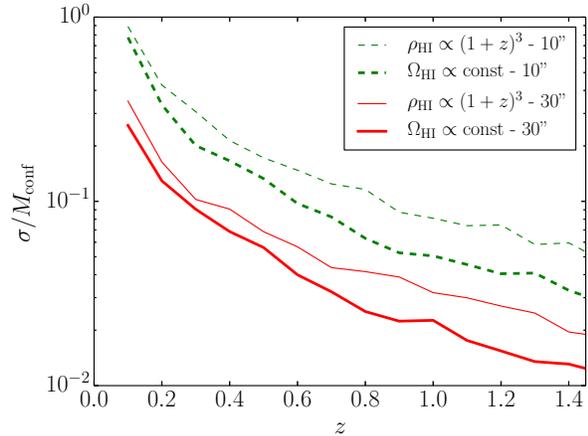}
\caption{The fractional uncertainty in the confused mass (standard deviation divided by mean value) estimated by simulating 100 stacks, each of 1,000 targets, at each redshift and beam size. The bold lines indicate those mocks which assume $\Omega_{\mathrm{HI}}$ is constant, and the standard weight lines are for a $\rho_{\mathrm{HI}} \propto (1+z)^3$ model. The solid (red) lines represent a beam size of 30 arcsec, and the dashed (green) lines a beam size of 10 arcsec. These are estimates of the 1-$\sigma$ fractional uncertainties in the  red (second from top) and green (second from bottom) solid lines in figures \ref{fig:conf_mass_const_Omega} \& \ref{fig:conf_mass_evo}.}
\label{fig:conf_mass_var}
\end{figure}
There is an error associated with our choice of the parametric forms used to fit both the 2D CF (equation \ref{eqn:2dcf}) and the mass-width function \citep[see ][]{Jones+2015}, as well as the exact ranges we chose to fit them over. As this is a single choice that involves the judgement of the individual performing the fit, it is very difficult to estimate a quantitative error for. Thus, rather than quoting an error we have chosen to a) demonstrate that the model we use is consistent with both the number counts and the observed rate of confusion between detections in the ALFALFA data set \citep{Jones+2015}, and b) present arguments (sections \ref{sec:red_evo} and \ref{sec:discuss}) that the two extremes which we adopt for any redshift evolution, likely bracket the true value. 

The above concerns aside, there is still another uncertainty that is important. Equation \ref{eqn:model} gives the confused mass that is present \emph{on average} in a stacked spectrum. As alluded to previously (section \ref{sec:mass_in_stack}) the variance of this quantity depends on the shape of the HIMF, and the more top-heavy it is, the higher the variance in $M_{\mathrm{conf}}$. In addition to the shape of the HIMF, the variance of $M_{\mathrm{conf}}$ also depends on the number of spectra being stacked, the angular size of the `cut outs', and the redshift of the stack. To estimate the scale of the variance we ran 100 realisations of stacks of 1,000 targets at redshifts 0.1 to 1.4 (in increments of 0.1), for two beam sizes, 10" and 30" (at $z=0$). Figure \ref{fig:conf_mass_var} shows the fractional uncertainties (standard deviation divided by the mean) in the confused mass calculated from these realisations. While the uncertainty for the mock stacks with a 30 arcsec beam quickly (by $z \sim 0.3$) drop to less that 10\%, for the stacks using a 10 arcsec beam the uncertainty starts off at almost 100\% and does not fall to 10\% until between a redshift of 0.5 and 1 (depending on the assumed evolution of $\Omega_{\mathrm{HI}}$). This indicates that accounting for confusion in a statistical way will be difficult for surveys with small beam sizes, as the variance in any individual stack will be so large. However, as is shown below confusion will turn out to be only a minor concern for surveys achieving beam sizes of 10 arcsec.

\section{Results}
\label{sec:results}

Before proceeding with predictions for upcoming surveys the CF model was tested against an existing study of HI stacking in a highly confused regime by \citet{Delhaize+2013}. In that paper HIPASS non-detections were stacked based on Two-Degree-Field Galaxy Redshift Survey (2dFGRS) positions and redshifts. The mean redshift of their sample was 0.029, and the stacked spectrum has a mass of $3 \times 10^{9}\,h_{70}^{-2}\,$\Msol \ between velocities $\pm300$ \kms. They also estimated that each source was confused with three others (on average), which increased the effective luminosity of the stacked sample by a factor of 2.5. Assuming a constant mass-light-ratio, this means that the contribution of confusion to the stack was approximately $1.8 \times 10^{9}\,h_{70}^{-2}\,$\Msol. A higher redshift sample of targeted follow-up was also stacked, giving a mean mass of $1.4 \times 10^{10}\,h_{70}^{-2}\,$\Msol \ at a mean redshift of 0.096, of which $1.1 \times 10^{10}\,h_{70}^{-2}\,$\Msol \ was estimated to be due to confusion.

Using our framework (and assuming constant $\Omega_{\mathrm{HI}}$) to estimate the confused mass in a stack at a redshift of 0.029 in HIPASS data returns a value of $1.9 \times 10^{9}\,$\Msol \ for a beam size of 15.5', and $3.3 \times 10^{9}\,$\Msol \ for a beam size of 21.9'. The Parkes telescope beam size is 15.5' for a wavelength of 21 cm, but the weighting used in \citet{Delhaize+2013} produces an effective beam size of 21.9' (and 21.2' for the higher $z$ sample). We quote results for both beam sizes as our model does not incorporate the beam profile weighting they assume. For the higher redshift sample we estimate a confused mass of between $1.3$ and $2.0 \times 10^{10}\,h_{70}^{-2}\,$\Msol \ for beam sizes 15.5' and 21.2' respectively. Both of these results appear approximately consistent, although the exact confidence is not possible to assess (see section \ref{sec:discuss}).

\begin{figure*}
\centering
\includegraphics[width=\textwidth]{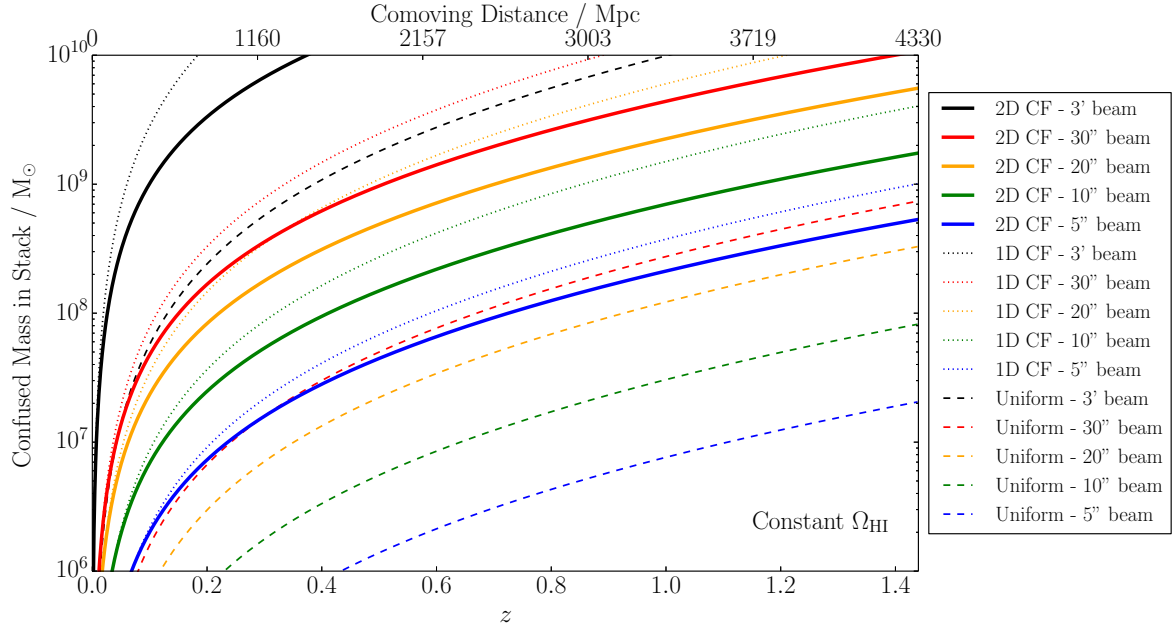}
\caption{The predicted average HI mass due to sources of confusion in a stacked spectrum as a function of the redshift (distance) of the stack, assuming $\Omega_{\mathrm{HI}}$ is fixed at its zero redshift value. The line styles indicate the method used to generate the estimate, with solid lines representing the 2D correlation function, dotted lines the projected (or 1D) correlation function, and dashed lines assume the Universe is uniform in HI. The blue, green, orange, red and black lines use beam sizes of 5, 10, 20, 30 arcsec, and 3 arcmin (at $z=0$) respectively, or equivalently smallest to largest beam going bottom to top.}
\label{fig:conf_mass_const_Omega}
\end{figure*}

\begin{figure*}
\centering
\includegraphics[width=\textwidth]{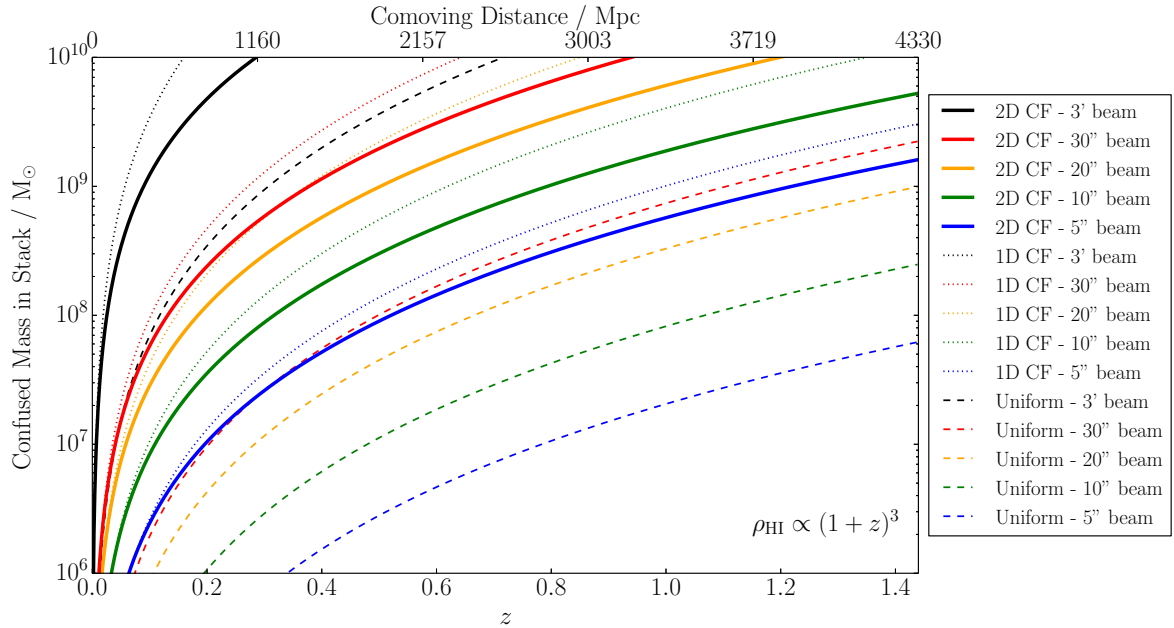}
\caption{Identical to the figure above except $\rho_{\mathrm{HI}}$ increases in proportion to $(1+z)^{3}$ from its zero redshift value.}
\label{fig:conf_mass_evo}
\end{figure*}

The confused mass present, on average, in a stack made from a generic survey at a given redshift, was estimated based on the integral of the 2D CF over the telescope beam and $\pm300$ \kms \ in redshift space (see section \ref{sec:mass_in_stack}). Figures \ref{fig:conf_mass_const_Omega} \& \ref{fig:conf_mass_evo} show the results for various telescope resolutions, each solid line represents a different angular resolution: 5, 10, 20, 30 arcsec and 3 arcmin (at $z=0$), from bottom to top. The dashed lines represent the confused mass that would be present if the Universe were perfectly uniform, and the faint dotted lines are the results obtained using the projected CF \citep{Papastergis+2013}, which removes the difference in the physical and velocity directions. The two figures are identical except that figure \ref{fig:conf_mass_const_Omega} assumes $\Omega_{\mathrm{HI}}$ does not change from its value at $z=0$, while figure \ref{fig:conf_mass_evo} assumes $\rho_{\mathrm{HI}}$ grows like $(1+z)^{3}$.


Below we outline the results relevant to each upcoming survey. Wherever a value is quoted for the constant $\Omega_{\mathrm{HI}}$ case, the $\rho_{\mathrm{HI}} \propto (1+z)^{3}$ value will immediately follow in parentheses (if different at the stated precision).

\subsection{CHILES}

CHILES has a resolution of 5 arcsec, so the solid blue (lowest) lines in figures \ref{fig:conf_mass_const_Omega} \& \ref{fig:conf_mass_evo} are the appropriate estimates of the confusion in stacked CHILES data. Even at the maximum redshift (0.45) the confused mass within one synthesised beam, and $\pm$300 \kms, will still only be $\sim10^{7}$ \Msol, indicating that CHILES will have no major concerns due to confusion when stacking sources. However, as CHILES will spatially resolve almost all the sources it detects, a more appropriate measure of the confused mass can be derived by choosing a constant physical scale for the `postage stamp' cut out of a galaxy in a stack. To be overly conservative we choose 100 kpc (diameter), which gives a confused mass of $1 \times 10^{8}$ \Msol \ at $z=0$, which increases to $1.6 \times 10^{8}$ \Msol \ ($3.1 \times 10^{8}$ \Msol) at $z=0.45$. In other words, CHILES would only encounter non-negligible amounts of confusion bias if very low mass objects (presumably at lower redshift) were to be stacked, which seems unlikely given the that CHILES is a pencil beam survey.

\subsection{LADUMA}

For LADUMA the angular size of the minimum synthesised beam is still not set, however as MeerKAT's maximum baseline will be smaller than the VLA's B-configuration baseline, here we assume LADUMA will have a resolution of 10 arcsec. This is represented by the solid green (second lowest) line in figures \ref{fig:conf_mass_const_Omega} \& \ref{fig:conf_mass_evo}. As mentioned above, in reality the confused mass is unlikely to ever drop much below $10^{8}$ \Msol \ even at low redshifts, as the physical size of the sources (rather than the size of the beam) will determine the `postage stamp' size. 

Again this indicates that LADUMA will be safe from the impact of confusion when stacking sources significantly more massive than $10^{8}$ \Msol, at least up to intermediate redshifts. By the outer edge of LADUMA's bandpass ($z=1.45$) the mass in confusion will have risen to $1.8 \times 10^{9}$ \Msol \ ($5.4 \times 10^{9}$ \Msol), potentially large enough to influence the stacking of $M_{*}$ galaxies.

However, if LADUMA were to be unable to achieve its intended synthesised beam size, then things would look quite different. The orange (third lowest) lines show the case for a 20 arcsec beam, which at the outermost redshift (1.45) would contain over $5 \times 10^{9}$ \Msol \ ($1.5 \times 10^{10}$ \Msol) of confused HI, and even by $z \sim 0.5$ would contain $5 \times 10^{8}$ \Msol \ ($1 \times 10^{9}$ \Msol). Preliminary estimates of LADUMA's detection capability (A. Baker, private communication) suggest that at $z \sim 0.5$ targets down to masses of $3 \times 10^{8}$ \Msol \ might be detectable via stacking, and by the outer edge of the survey this will have increased to $3 \times 10^{9}$ \Msol. In both cases, if LADUMA were to have a beam size of 20 arcsec rather than 10, then these stacks would contain more mass in confused HI than in the target objects. While this may not prevent progress via stacking, it would add a strong additional bias and a new level of complexity to the process that would require careful consideration, compared to if the survey achieves its target resolution.

\subsection{DINGO UDEEP}

Similarly to CHILES, if ASKAP is able to achieve 10 arcsec resolution then the stacking capabilities of DINGO UDEEP will be limited by the physical size of objects, rather than the survey's angular resolution, throughout most of its redshift range (0.1-0.43). Whereas, if only a 30 arcsec resolution can be achieved then, as the red (second highest) line in figures \ref{fig:conf_mass_const_Omega} \& \ref{fig:conf_mass_evo} shows, the confused mass will soon rise well above $10^{8}$ \Msol, complicating the interpretation of any stacks of objects of comparable mass. Although stacking of objects above $10^{9}$ \Msol \ should still be relatively unimpeded, as the confused mass does not reach $10^{9}$ \Msol \ until $z \sim 0.4$ and, as will be discussed in section \ref{sec:mit_strat}, the confusion signal can be effectively removed until it becomes comparable to the target signal.

Using the \citet{Jones+2015} expression for a general survey detection limit and assuming an order of magnitude improvement from stacking, we estimate that DINGO UDEEP will be capable of detecting an object with an HI mass of $3 \times 10^{8}$ \Msol \ via stacking at $z=0.2$, but at that redshift the predicted confused mass is $1.7 \times 10^{8}$ \Msol \ ($2.4 \times 10^{8}$ \Msol) for a 30 arcsec beam. At $z=0.4$ the situation is slightly worse, with the confused mass becoming $6.2 \times 10^{8}$ \Msol \ ($1.1 \times 10^{9}$ \Msol) and the mass detectable via stacking being $1 \times 10^{9}$ \Msol.

\subsection{FAST}

Unlike the other telescopes discussed here FAST is a single dish, and thus will have a much poorer resolution. The black solid (highest) line in figures \ref{fig:conf_mass_const_Omega} \& \ref{fig:conf_mass_evo} shows the expected confused mass for a FAST based survey, which rises above $10^{9}$ \Msol \ by a redshift of $\sim$0.1 and by 0.4 (0.3) even the most HI massive galaxies will be severely impacted by confusion. FAST's vast collecting area will mean it might be capable of directly detecting HI galaxies in a survey out to $z=0.2$ or greater, and would certainly by capable of doing so via stacking, but regardless of how these sources might be detected they will still be subject to considerable bias from confusion.

\section{Discussion}
\label{sec:discuss}

The approximate agreement shown between the estimates of the confused mass from stacks of Parkes data \citep{Delhaize+2013} and our model is an encouraging validation. However, both our predictions are somewhat higher than the estimates from that paper. The significance of this is difficult to assess as the values quoted from \citet{Delhaize+2013} are not given with errors at the relevant stage of their calculation. The simplest potential explanation might be that this is variance between the average value and two particular examples, however using similar multiple realisations of mocks stacks to those in section \ref{sec:mod_var} it is clear that this cannot be the explanation, as we measure only a standard deviation of approximately a percent between equivalent simulated stacks.

If this offset is real then the reason for it is uncertain; one possibility is that this model uses the HI auto-correlation function, whereas the HI-optical cross-correlation function might be the most appropriate. As shown in \citet{Papastergis+2013} the correlation function of SDSS blue galaxies is almost indistinguishable from that of an HI population, but HI-rich galaxies are much less likely to be found in regions with high densities of red galaxies. Therefore, an input sample that contains any red galaxies will have less confused HI mass around those targets than would targets based on an HI selected sample. Another possible explanation is that the assumption of a constant HI mass to light ratio across the target and confused sources might not be valid at the level of the discrepancy.

Assuming that the upcoming interferometric HI surveys can achieve their desired beam sizes they should have minimal amounts of confusion when making stacks throughout most of their bandpass ranges. However, due to its beam size, confusion is considerably more worrisome for FAST. \citet{Duffy+2008} estimated the contribution of confusion to a FAST survey and found that even for very long integration times (over 15 hours) it would not be a concern until beyond a redshift of 0.5. The dashed black (highest) line in figure \ref{fig:conf_mass_evo} shows the confused mass calculated assuming a uniform universe for a FAST sized beam (3 arcmin). This is equivalent to how the confused mass was defined by \citet{Duffy+2008}, but our value of $\Omega_{\mathrm{HI}}$ is 16 percent larger. As can be seen here the inclusion of the CF (solid black line), compared to assuming a uniform background, increases the confused mass by more than an order of magnitude (over the relevant redshift range). This will severely limit FAST's ability to probe HI galaxies much beyond $z=0.1$, which reiterates the conclusion of \citet{Jones+2015}, that future blind HI surveys with single dish telescopes should focus on the nearby universe where their larger beam sizes are a strength rather than a hindrance.
\begin{figure*}
\centering
\includegraphics[width=\columnwidth]{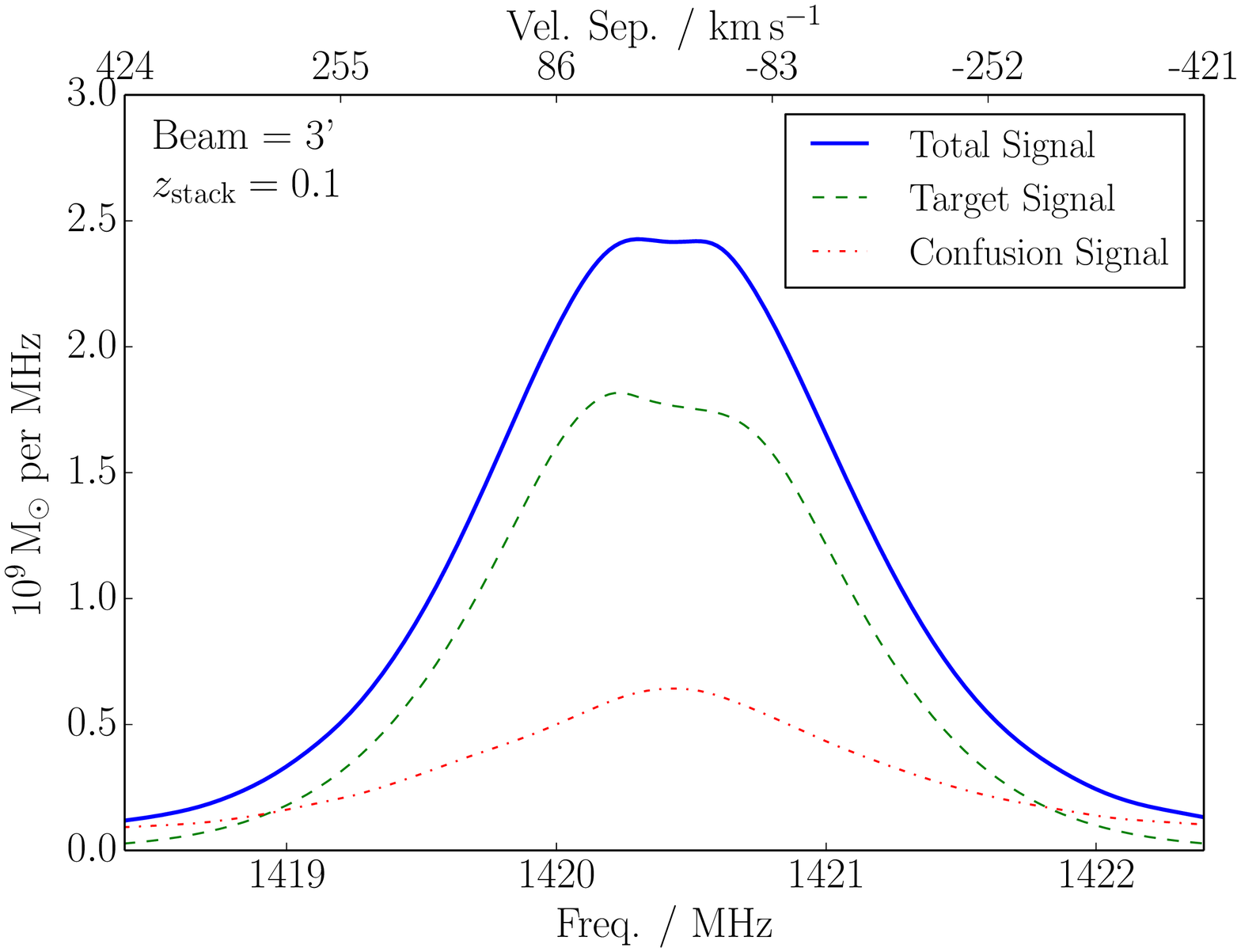}
\includegraphics[width=\columnwidth]{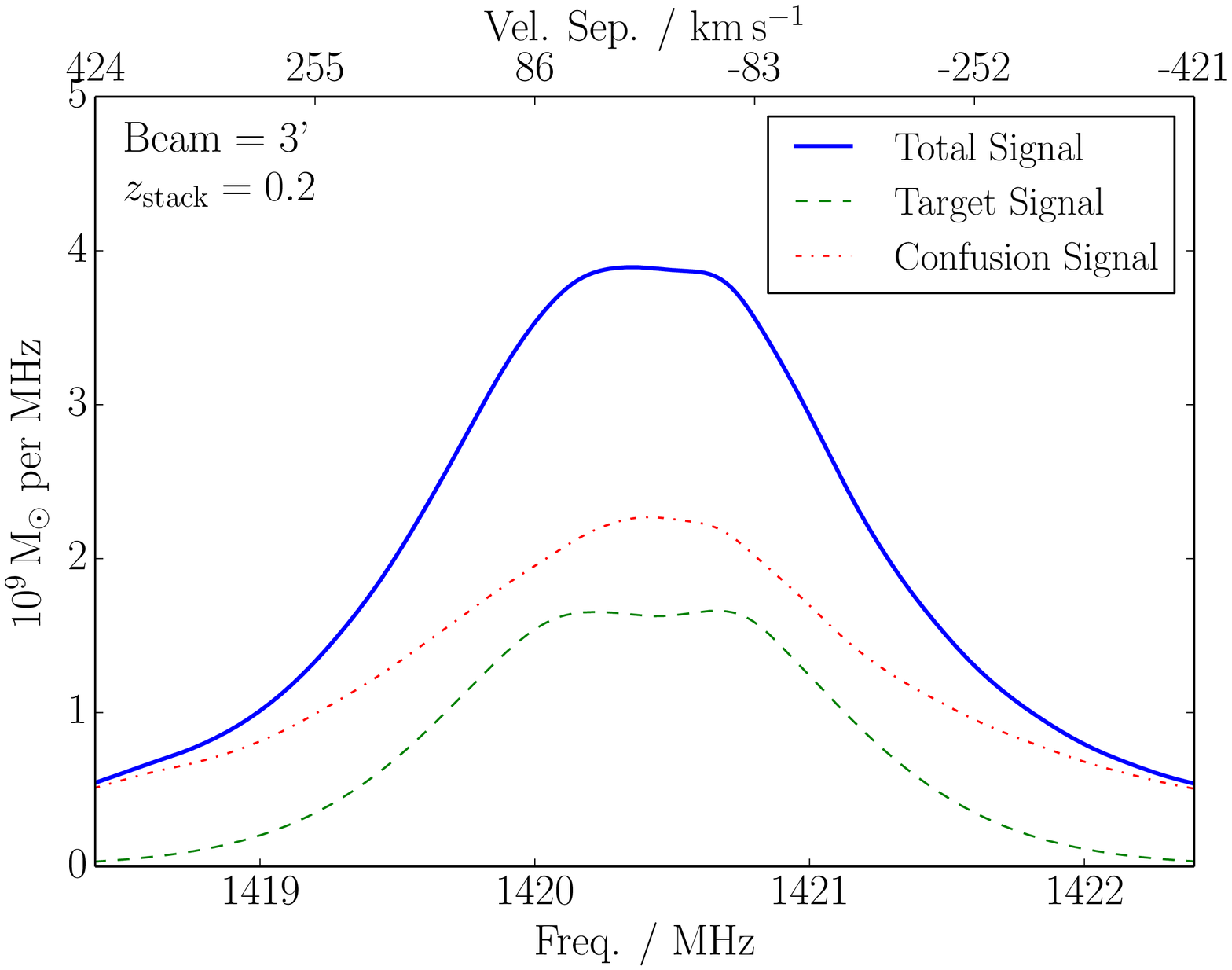}
\caption{Simulated noiseless stacks of 1,000 galaxies of HI mass $3 \times 10^{9}$ \Msol, showing the contributions from the target galaxies (dashed green lines) and from confusion (red dash-dot lines). Both plots assume a zero redshift beam size of 3 arcmin (as expected for FAST). The left plot is for a stack at $z=0.1$, and the right at $z=0.2$. Both assume $\rho_{\mathrm{HI}} \propto (1+z)^{3}$.}
\label{fig:FAST_stack}
\end{figure*}

To show how confusion may affect a stack of data from a FAST survey we have simulated two stacks of galaxies with target objects of $3 \times 10^{9}$ \Msol, at redshifts 0.1 and 0.2 (see figure \ref{fig:FAST_stack}). If the total signal is (incorrectly) assumed to be made up of two Gaussian components, a broad one due to confusion and a narrow one due to the target signal, the mean target masses are found to be 3.3 and $4.4 \times 10^{9}$ \Msol \ respectively at $z = 0.1$ and 0.2. The excess signal that is incorporated into the narrow Gaussian component originates from the fact that the confusion profile is itself a double Gaussian, and is therefore not adequately subtracted by the broad component alone. In fact the overestimation would be worse, but some of the target signal is clipped (by the $\pm 300$ \kms \ boundary), and some is incorporated into the broad Gaussian along with the confusion signal.

A major uncertainty in our predictions is redshift evolution, which due to the current lack of data is inadequately modelled. We argued in section \ref{sec:red_evo} that the two cases presented for the evolution of HI density likely bracket the true evolution in that quantity, however the impact of the change in the HI CF is more difficult assess. \cite{Hartley+2010} find that the correlation length of blue galaxies in the UKIDSS Ultra Deep Survey increases by approximately a factor of 2 going from $z=0$ to 1.5. At $z=0$ blue galaxies and HI-rich galaxies are proxies for each other. Therefore, it is reasonable to assume that the HI CF would also be raised with increasing redshift, meaning the curves shown in figure \ref{fig:conf_mass_const_Omega} would represent lower limits on the confused mass in stacks.

The two models of the evolution of $\rho_{\mathrm{HI}}$ unsurprisingly give similar results at low redshift, but start to diverge at larger redshift, leaving LADUMA with the most uncertain measure of confused mass. The shape of the confused mass versus redshift curve for the constant $\Omega_{\mathrm{HI}}$ model (figure \ref{fig:conf_mass_const_Omega}) is qualitatively similar to the shape of a model detection limit for an HI survey. \citet{Fabello+2011} found that an order of magnitude below the detection limit is the most that could be gained by stacking, before non-Gaussian noise became dominant (although \citet{Delhaize+2013} indicates that deeper stacks might be possible with very well characterised noise). Therefore, assuming that at all redshifts there are sufficient stacking targets available that are approximately an order of magnitude below the detection limit, we arrive at the somewhat counter intuitive result that the ratio of the mean mass of these targets to the confused mass in their stack, is almost independent of redshift.\footnote{Note that this may appear to be in conflict with the \citet{Delhaize+2013} experiment, however that is because their two datasets have very different integration times, allowing them to probe lower masses than would otherwise be possible in their higher redshift sample, and thus making the stack more confused.} It should be noted however that this will break down at the lowest redshifts because, as stated previously, in practice the physical size of galaxies will prevent the confused mass ever dropping much below $10^{8}$ \Msol. In the case where $\rho_{\mathrm{HI}}$ increases with the Universe's decreasing volume (figure \ref{fig:conf_mass_evo}) the confused mass rises much more steeply with redshift, producing much more severe confusion at higher $z$. While this might seem like the most conservative model to use, the currently available data (\citet{Rhee+2013} and references within) indicate that $\rho_{\mathrm{HI}}$ does not increase this quickly with redshift.
\begin{figure*}
\centering
\includegraphics[width=\columnwidth]{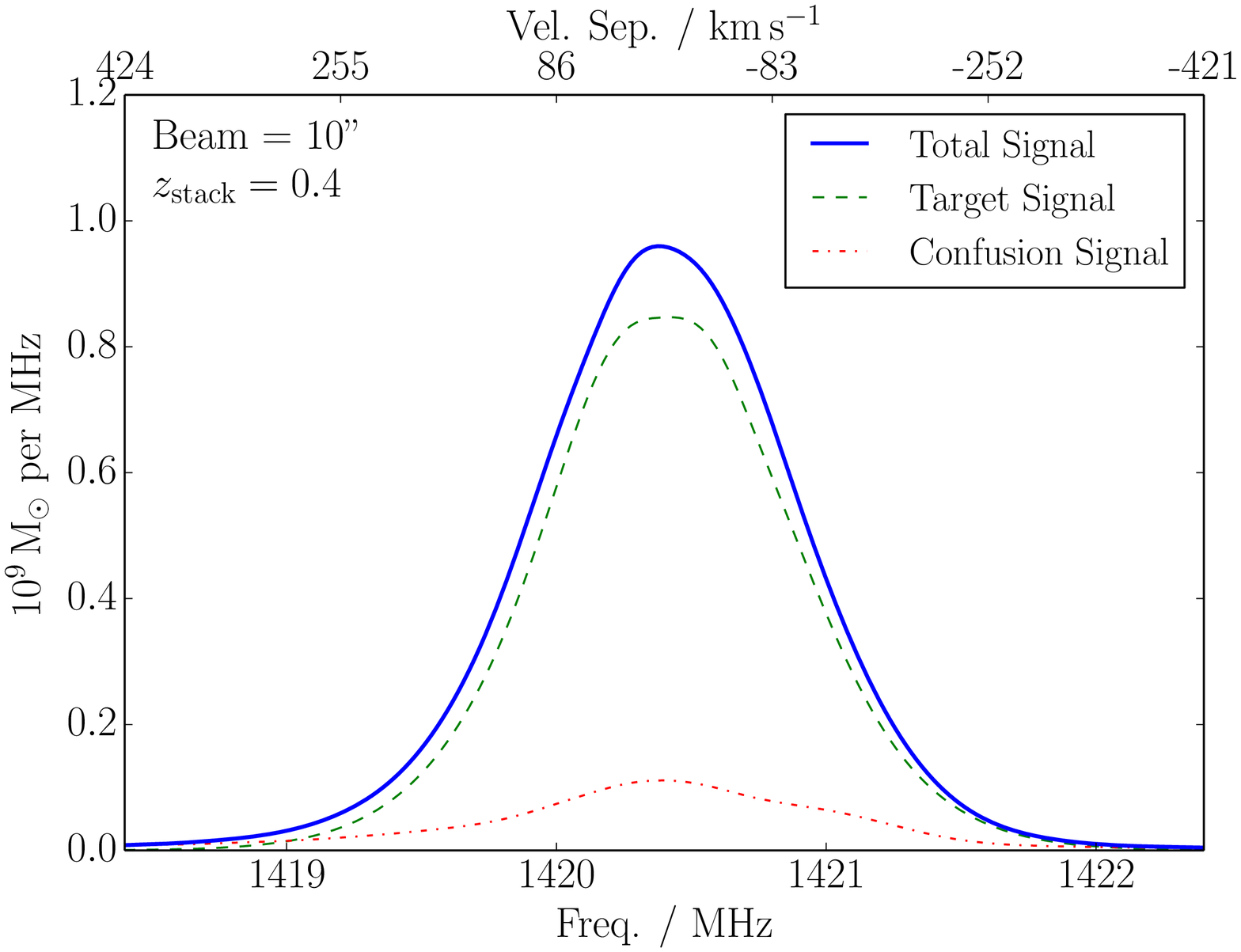}
\includegraphics[width=\columnwidth]{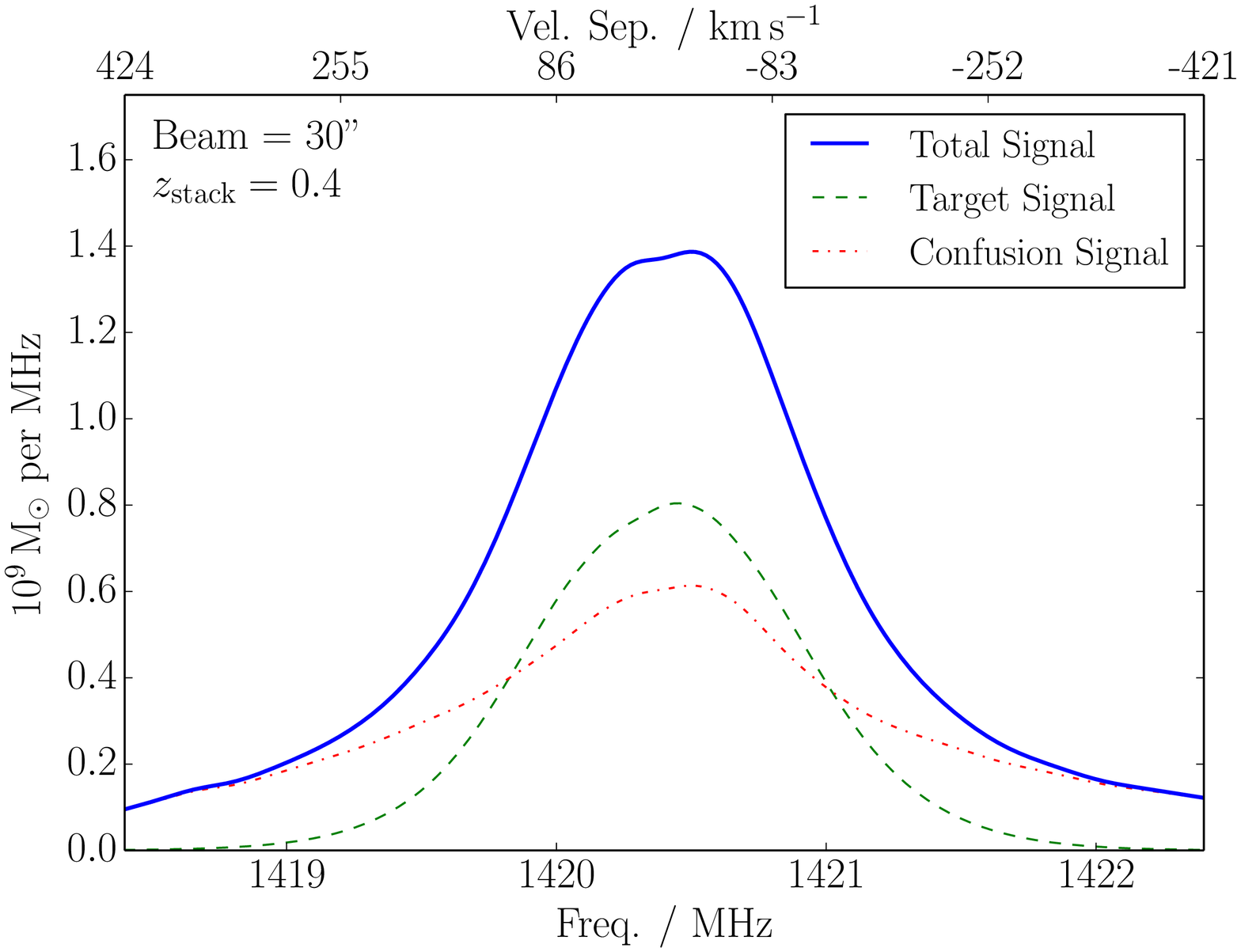}
\caption{Simulated noiseless stacks of 1,000 galaxies of HI mass $10^{9}$ \Msol, showing the contributions from the target galaxies (dashed green lines) and from confusion (red dash-dot lines). Both plots are for a stack at $z=0.4$ and assume that $\rho_{\mathrm{HI}} \propto (1+z)^{3}$, but the left has a zero redshift beam size of 10 arcsec, while the right has a 30 arcsec beam. }
\label{fig:DINGO_stack}
\end{figure*}

Regardless of which evolution model is assumed to be correct, the results show that for surveys like LADUMA and DINGO UDEEP, where the synthesised beam size is not yet fixed, there is much to be gained in terms of the stacking performance by pushing to a lower beam size (in this case 10 arcsec). The difference in confused mass between a beam size of 10 and 30 arcsec is approximately an order of magnitude. For DINGO UDEEP a 30 arcsec beam would mean that a large fraction of the mass in stacks (probing the lowest possible HI masses) will be contributed by confusion, at all redshifts; whereas with a 10 arcsec beam the contribution would be almost negligible. For LADUMA there is little option but to use a $\sim$10 arcsec beam if stacking is going to be a viable option. Even with a 20 arcsec beam the smallest masses that are in theory detectable via stacking would likely always be below the level of the confused mass, but with a 10 arcsec beam this would not be the case until the very largest redshifts.

Figure \ref{fig:DINGO_stack} shows the contributions of confusion in two simulated stacks at approximately the outer edge of DINGO UDEEP's bandpass ($z=0.4$) for beam sizes of 10 and 30 arcsec (at $z=0$). The target galaxies have HI masses of $10^{9}$ \Msol, the lowest that will likely be detectable via stacking with this survey at $z=0.4$. While the 30 arcsec beam introduces $1.1 \times 10^{9}$ \Msol \ of confusion, the 10 arcsec beam only introduces $1.5 \times 10^{8}$ \Msol. In this case naively splitting the resulting total signal into two Gaussian components gives a mean target mass of 1.3 and $1.0 \times 10^{9}$ \Msol, for the 30 and 10 arcsec beams respectively.

For regimes where the confused mass in a stack is comparable to the anticipated mass of the targets, the spectral profile calculated in section \ref{sec:spec_prof} indicates that caution must be used. The profile of confusion alone appears to be well fit by a double Gaussian, where the two components arise from the width of the velocity space CF and the distribution of redshift uncertainties in the input catalogue. This is precisely the profile that might be expected from a stack detection with a small amount of confusion, a narrow Gaussian (presumed from the target objects) superimposed on a broader Gaussian (presumed to be from confusion). Thus, in a severe case it is possible that confusion alone could be misidentified as a detection and confusion. In a more moderate case, where there is a real detection, it is desirable to minimise the amount of confused mass and to understand how much it still contributes to the final stack. Strategies to accomplish this are discussed below.

\subsection{Mitigation Strategies}
\label{sec:mit_strat}

In any stacking experiment where a significant contribution from confusion is anticipated (not limited to the surveys discussed here), there are two approaches that can be taken to improve the outcome: either strategies to remove confused mass can be implemented, or the amount of the final signal that is contributed by confusion can be estimated.

As a first approximation the model presented in this paper can be used to predict how much confusion there is in a stack, however there are a number of situations where this might give a poor estimate. For example, a stack with a small number of targets, at high redshift, or with an input catalogue of galaxies not selected for HI content. In such cases other strategies might be necessary. One approach could be to explore the properties of such stacks in a simulation, another is to attempt to mitigate the impact of confusion when extracting the final parameters from a stack, which is the approach we discuss below.

\subsubsection{Double Gaussian Decomposition}

As figure \ref{fig:spec_prof} shows, a large fraction of the signal from confusion is expected to be in a broad Gaussian component, whereas most of the target emission should be in a narrow component. Although there is also a narrow component to the confusion profile, removing the broad component will help to alleviate much of the confusion. 

This approach was tested by simulating the confusion in a stack using representative HI line profile shapes \citep{Saintonge+2007}, positions from the CF, and assuming the $z=0$ value of HI density \citep{Martin+2010}. The narrow Gaussian component of the total profile was found to reproduce the mean target mass well (within $\sim$10\%) in the cases where the confused mass was less that about 2/3 of the target mass, although results were marginally worse for more massive, broader targets. Presumably the portion of the target signal that is excluded from the narrow Gaussian is approximately made up for by the inclusion of some of the narrow component of confusion. However, when the confused mass becomes almost as large as the target mass, the narrow Gaussian integral begins to diverge from the mean target mass.

Thus, this straightforward method is very successful for stacks with low levels of confusion, but cannot adequately separate target signal and confusion when the confusion is more severe.

\subsubsection{Beam Weighting}
\label{sec:beam_weight}

In the regime where the telescope beam (or synthesised beam) is considerably larger than the target source, the weighting of the pixels can be tapered away from the target. This will have little impact on the target flux (presumably concentrated in the central pixel), but will give lower weight to the surrounding confusion signal.

This approach was tested using mock stacks, as before. For stacks with physical beam sizes of 100-600 kpc, assuming 4 pixels across a beam width and a Gaussian weighting scheme, the confused mass was reduced by approximately 25-30\% compared to a uniform weighting. However, for larger beam sizes there are diminishing returns as in addition to the target, many of the confused objects also lie within the central pixel.

\subsubsection{Inclusion and Exclusion of Confused Targets}

Possibly the most obvious solution to confusion is to simply excluded the stacking targets that are likely to be heavily confused. Most of the HI mass in the Universe is contained in $M_{*}$ systems, which are likely to be visible in the optical input catalogue. Targets that are in close proximity to $M_{*}$ galaxies (provided they have optical redshifts) could in principle be removed from the input catalogue. As most of the HI mass is contained in these galaxies, this would remove most of the confused mass from the stack.

This approach has some promise for the cases where low mass galaxies are being stacked at low redshift and $M_{*}$ galaxies are uncommon, but for higher redshifts where the beam sizes become larger, many targets are confused, often multiple times \citep{Delhaize+2013,Jones+2015}. Thus, it becomes impractical to remove them.

The approach taken by \citet{Delhaize+2013} was to include such targets, but to note the presence of likely sources of confusion. By assuming a constant HI mass to light ratio they were able to estimate the fraction of the stack mass that was contributed by confusion. As shown in section \ref{sec:results} our results are roughly consistent with their findings. This procedure could be taken further by using HI scaling relations with stellar mass or disc size to improve the estimate of the confused galaxy masses \citep{Toribio+2011,Huang+2012}.

\subsubsection{Exclusion in Velocity Space}

Weighting the beam cuts confusion by eliminating sources spatially, but this can also be done in velocity space. \citet{Fabello+2011} used the Tully-Fisher relation (TFR) to remove the section of the spectrum containing the intended target, in order to estimate the rms noise in the rest of the spectrum. The same method could be used to stack just the region of the spectrum that is likely to contain emission from the target galaxy, thereby removing additional sources in front or behind the target that would otherwise contribute to a stack made with a conservative $\pm 300$ \kms \ cut.

This method was simulated as before, but with each contributing spectrum cut off at $\pm (W_{TF}/2 + \sigma_{\mathrm{input}})$ away from the target redshift. Where $W_{TF}$ is the target's simulated velocity width ($W_{50}$) with 0.2 dex of scatter introduced (designed to emulate the TFR), and $\sigma_{\mathrm{input}}$ is the standard deviation of the redshift uncertainty in the input catalogue (35 \kms). This gave approximately a 60\% reduction in confused mass when stacking targets in the mass range $10^{8}$ - $10^{9}$ \Msol, and a 45\% reduction for targets in the range $10^{9}$ - $10^{10}$. However, it also typically removed 30-35\% of the flux from to the target objects, with the higher mass stack more effected.

\section{Conclusions}

We created a model to predict the average amount of HI mass contributed by confused sources to a stack from a generic survey. The analytic expression of our model (equation \ref{eqn:model}) is derived in the general case, allowing for different beam sizes, velocity ranges, HI background densities or fits to the CF to be used to make quick estimates of the amount of confusion in any HI survey. This model, based on the ALFALFA correlation function, shows agreement with estimates of the confusion present in stacks of Parkes data \citep{Delhaize+2013}, and predicts approximately an order of magnitude more confused HI than found from assuming a uniform universe \citep{Duffy+2008}. 

The largest uncertainty in the predictions comes from our relative ignorance of the redshift evolution of HI-rich galaxies. However, we argued that the true values likely fall between the two idealised cases presented here, and that the smaller of the two is in fact a lower limit.

The results for upcoming SKA precursors surveys, like LADUMA and DINGO UDEEP, reveal that it would be highly advantageous if these surveys could achieve their initially intended resolutions (10 arcsec), as any resolution substantially poorer than this would lead to stacks that are dominated by confusion, rather than their target objects.

Confusion was the most concerning for FAST; its larger (single dish) beam size results in the mass in confusion rapidly overtaking even that of $M_{*}$ galaxies, as redshift increases. This will prevent a FAST based blind HI survey from probing individual galaxies much beyond $z=0.1$ with either stacking or direct detections. Similarly to the findings of our previous work \citep{Jones+2015} this indicates that single dish telescopes should focus their HI galaxy studies on the local Universe. 

When simulating stacks with a large component of confusion we had limited success in implementing mitigation strategies. Weighting pixels in a Gaussian pattern reduced the confused mass by about 30\%, but is only suitable when one pixel is larger than the angular extent of the targets. Using the TFR to exclude regions of the spectrum beyond the target's emission was even more successful at removing unwanted confusion, however it also removed around 30\% of the target emission. Simply decomposing the total spectrum into broad and narrow Gaussian components was very successful at estimating the mean target mass with even moderate levels of confusion, despite it not being an accurate model of the profile shape of targets combined with confusion. However, when the confused mass approached that of the targets, the results began to diverge from the true values. Thus in the event of of heavily confused stack, the best approach will likely be not to try to exclude sources of confusion, but to use optical data or simulations to model and account for their HI properties.

\section*{Acknowledgements}

The authors acknowledge the work of the entire ALFALFA collaboration in observing, flagging, and extracting the catalogue of galaxies that this paper makes use of. The ALFALFA team at Cornell is supported by NSF grants AST-0607007 and AST-1107390 to RG and MPH and by grants from the Brinson Foundation. EP is supported by a NOVA postdoctoral fellowship at the Kapteyn Institute. MGJ would like to thank both Kelley Hess and Andrew Baker for helpful discussions concerning CHILES and LADUMA.

\bibliography{confusion_refs}

\end{document}